\documentclass{article}
\usepackage{spconf,amsmath,graphicx}
\usepackage{bbold}
\usepackage{xcolor}
\usepackage{bbm}
\usepackage{amssymb}
\usepackage{graphicx} 
\usepackage{xcolor}
\usepackage{enumitem}
\usepackage{xcolor}
\usepackage{multirow}
\title{
Sparse signal reconstruction 
for overdispersed low-photon count biomedical imaging
using $\ell_p$ total variation }
\name{Yu Lu and Roummel F. Marcia \thanks{{This work is partly supported by National Science Foundation grants DMS 1840265 and IIS 1741490.}}}
\address{Department of Applied Mathematics, University of California, Merced, Merced, CA, 95343 USA}
\begin{document}
%\ninept
%
\maketitle
\begin{abstract}
The negative binomial model, which generalizes the Poisson distribution model, can be found in applications involving low-photon signal recovery, including medical imaging. Recent studies have explored several regularization terms for the negative binomial model, such as the $\ell_p$ quasi-norm with $0 < p < 1$, $\ell_1$ norm, and the total variation (TV) quasi-seminorm for promoting sparsity in signal recovery.  These penalty terms have been shown to improve image reconstruction outcomes.
In this paper, we investigate the $\ell_p$ quasi-seminorm, both isotropic and anisotropic $\ell_p$ TV quasi-seminorms, within the framework of the negative binomial statistical model. 
This problem can be formulated as an optimization problem, which we solve using a gradient-based approach. We present comparisons between the negative binomial and Poisson statistical models using the $\ell_p$ TV quasi-seminorm as well as common penalty terms. Our experimental results highlight the efficacy of the proposed method.
\end{abstract}
\begin{keywords}
Negative binomial distribution, low-count imaging, nonconvex optimization, $\ell_p$ total variation quasi-seminorm.
\end{keywords}
\section{Introduction}
\label{sec:intro}
In numerous photon-limited signal processing applications, such as medical imaging \cite{willett2003platelets}, astronomy \cite{lanteri2005restoration}, and network traffic analysis \cite{Slimane95}, the Poisson model \cite{HarMW12} has been widely used for image reconstruction. 
However, the Poisson model assumes the data has the same mean and variance \cite{gardner1995regression}.
The negative binomial (NB) model is a more generalized statistical distribution, to which the Poisson model reduces under certain assumptions.
The negative binomial model has been used for low-photon signal reconstruction \cite{LuMBounds23} and matrix factorization \cite{gouvert2020negative}. Here, we explore the application of the negative binomial model for sparse signal recovery.

The negative binomial probability mass function (PMF) is given by
\begin{align*}
P(y| r, p) = \binom{y+r-1}{y}(1-\beta)^y {\beta}^r,
\end{align*}
where $y$ and $r$ represent the counts of successful and failure events, respectively. Here, $\beta$ denotes the failure probability. In general, the negative binomial probability can be thought of as the probability of observing $y$ successful events occurring before the $r^{\text{th}}$ failure, where each event is an independent and identically distributed Bernoulli trial with failure occurring probability $\beta$. 
The expectation and variance of the negative binomial PMF, denoted by $\mu$ and $\sigma^2$ respectively, are given by $\mu = r(1-\beta)/\beta$ and $\sigma^2 = r(1-\beta)/\beta^2 = \mu + \mu^2 / r$  \cite{degroot2012probability}. Given that the negative binomial distribution does not require identical mean and variance values. However, if we let $r$ go to infinity, the variance will approach the expectation, which shows Poisson distribution is a limiting case of negative binomial distribution.

For optimization methods in sparse signal recovery application, sparsity-promoting penalties can improve reconstruction accuracy.
Both the $\ell_p$ quasi-norm (for $0 < p < 1$) and the total variation (TV) quasi-seminorm have been evaluated in the contexts of both the Poisson \cite{orkusyan2016analysis, sawatzky2009total} and negative binomial models \cite{LuM23, LuMpNorm23}. The $\ell_p$ TV quasi-norm \cite{liu2018infrared} has been used successfully for the Poisson model \cite{adhikari2015p} for signal recovery.
This paper investigates the application of the $\ell_p$ TV quasi-norm in the framework of the negative binomial model.

%%%%%%%%%%%%%%%%%%%%%%%%%%%%%%%%%%%%%%%%%%%%%%%%%%%%%%%%%%%%%%%%%%
\section{Problem Formulation}
\label{sec:format}

\begin{figure*}[t]
\centering
\begin{tabular}{ccccccc}
\hspace{-.26cm} 
\includegraphics[width=3.4cm]{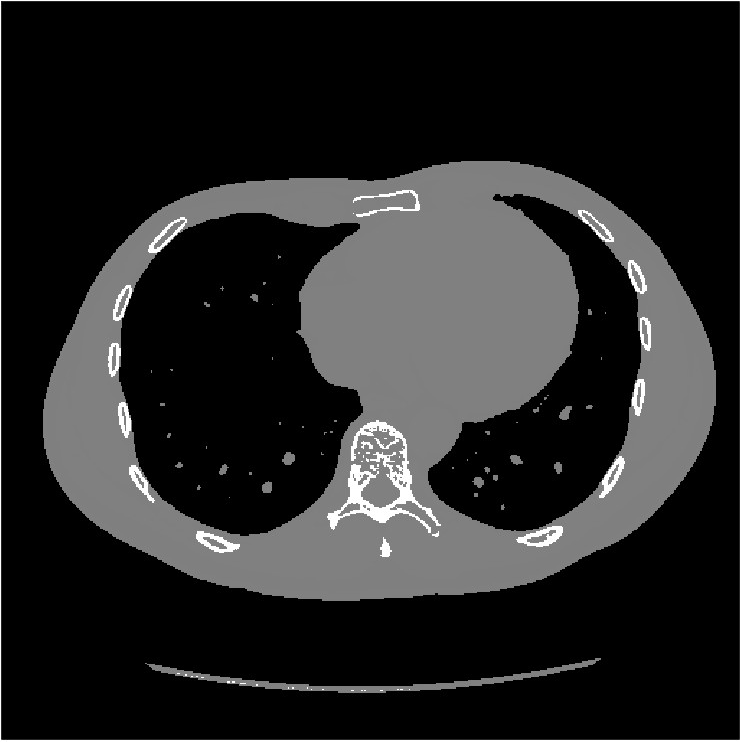} 
\hspace{-.22cm} 
&
\hspace{-.22cm} 
\includegraphics[width=3.4cm]{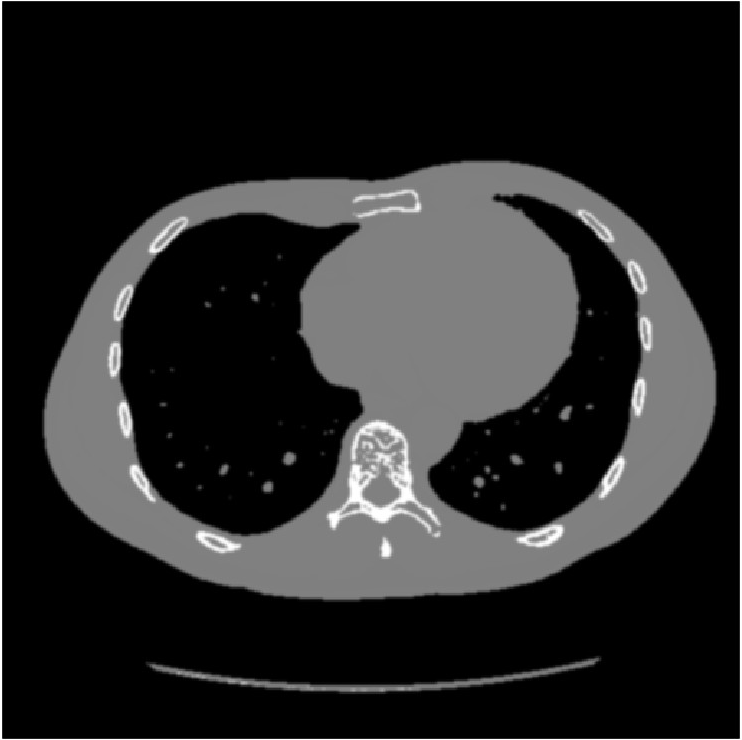} 
\hspace{-.22cm}  
&
\hspace{-.26cm} 
\includegraphics[width=3.4cm]{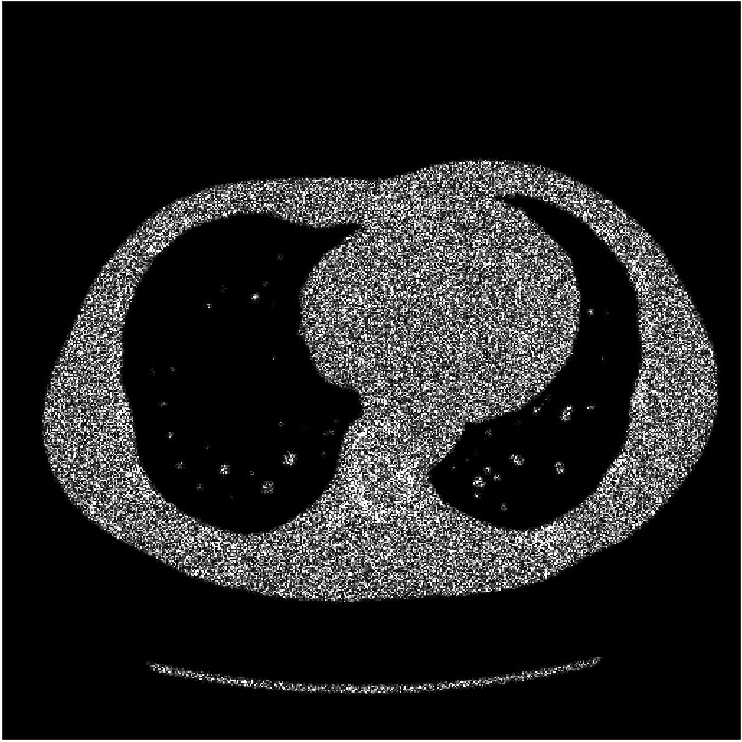} 
\hspace{-.22cm} 
&
\hspace{-.22cm} 
\includegraphics[width=3.4cm]{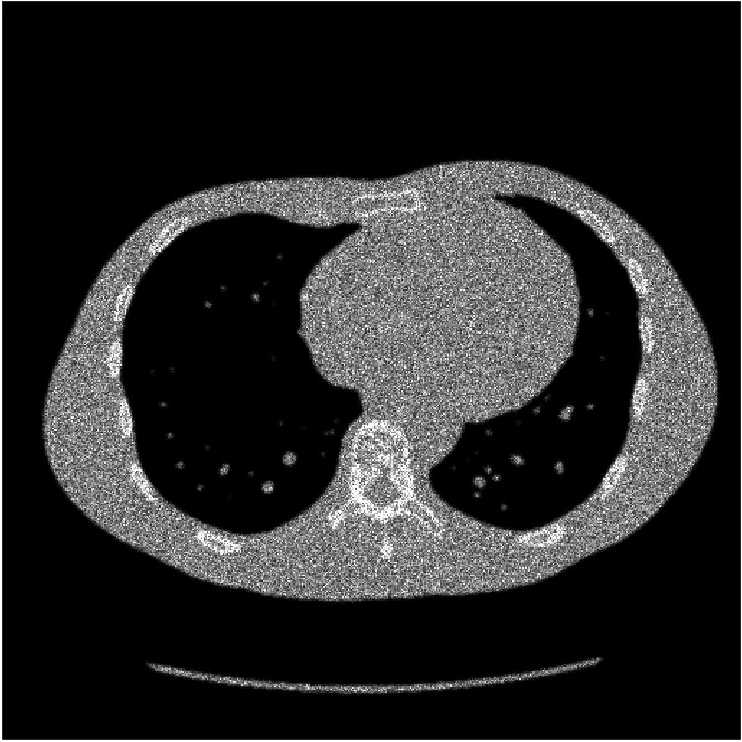} 
\hspace{-.22cm}  
&
\hspace{-.22cm} 
\includegraphics[width=3.4cm]{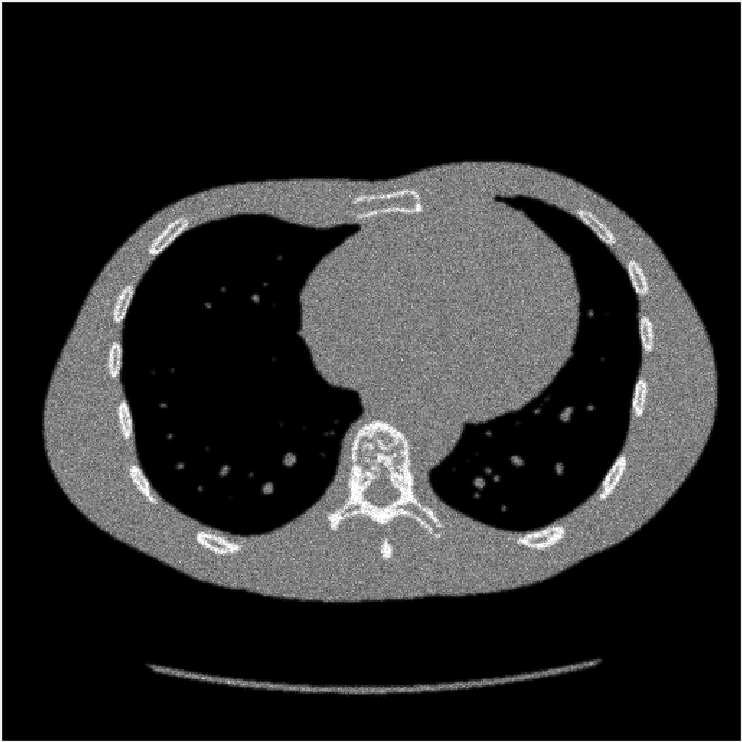} 
\hspace{-.22cm}  
\\
(a) $f^*$ & (b) $Af^*$ & (c) $y_\text{r=1}$  
& (d) $y_\text{r=10}$ & (e) $y_\text{r=1000}$ 
\end{tabular}

\caption{Example of observation model.  
(a) The true image $f^*$.
(b) The expected observation at the detector stage.  Here, the measurement operator $A$ is a Gaussian blur.  
(c) Observed measurement $y_{r=1}$ drawn from a negative binomial  (NB)  distribution with dispersion parameter $r = 1$.
(d) Observed measurement $y_{r=10}$ drawn from a NB  distribution with $r = 10$.
(e) Observed measurement $y_{1000}$ drawn from a NB  distribution with $r = 1000
$.}
\label{fig:exp}
\end{figure*}

Under the assumption of low-photon signal models, the observed vector of data $y$ can be drawn from negative binomial distribution, which can be expressed as
$$ 
y \sim \hbox{NB}(r, \beta).
$$
Replacing the probability $\beta$, the observation model becomes
\begin{align*}
y_i \sim \text{NB} \left (r, \frac{r}{r+(Af^*)_{i}} \right ).
\end{align*}
Here, $f^* \in \mathbb{R}^n_+$ denotes the true signal or image, while $A \in \mathbb{R}^{m \times n}_+$ represents the sensing matrix that projects the scene $f^*$ onto an expected measurement set $\mu$, given by $Af^*$. We denote the vector of noisy observation by $y \in \mathbb{R}^{m}_+$. This leads to the negative binomial statistical model 
\begin{align}\label{eq:pop}
     P( y | f ) 
     \!=\!  \prod_{i=1}^{m} \!
     \binom{ r\!+\!y_i\!-\!1  }{y_i} \!\! 
     \left ( \! \frac{r}{r\!+\!(Af)_i} \! \right )^{\!\! r}
     \!\!\!  
     \left ( \! \frac{(Af)_i}{r\!+\!(Af)_i} \! \right )^{\!\! y_i}\!\!\!.
\end{align}
See \cite{degroot2012probability} for further details.

To optimize the probability of observing $y$ in \eqref{eq:pop}, we apply the maximum likelihood principle and minimize the corresponding negative binomial log-likelihood function, which is given by
\begin{align}\label{eq:opt}
F(f) = \sum_{i=1}^m (r+y_i) \log(r+(Af)_i) - y_i \log((Af)_i).
\end{align}

In our application of interest, the pixel intensities of the images are piece-wise constant. Consequently, the variations in these intensities are sparse. We exploit this inherent property of the signal to improve the accuracy of signal recovery. In particular, we investigate the use of the $\ell_p$ TV quasi-seminorm penalty with $0 < p < 1$ to promote sparsity in our solution. 
The completed formulation of our problem can be expressed as
\begin{align}\label{eq:NBopt}
    f^{*} \ = \  &  \ \underset{f \in \mathbb{R}^n_+}{\text{arg min}} \quad \ \ \ \Phi(f) \equiv F(f) + {\tau} \| f \|_{TV}^p \nonumber \\
    &\ \hbox{subject to} \quad  0 \le f,
\end{align}
where $\tau > 0$ is a regularization parameter. There are generally two choices for the TV quasi-seminorm: the anisotropic
\begin{align*}
\displaystyle \Vert f \Vert^{p}_{TV^{(\text{A})}} = \sum_{l=1}^{m-1}\sum_{k=1}^{n}\vert f_{l,k}-f_{l+1,k}\vert^{p}+\sum_{l=1}^{m}\sum_{k=1}^{n-1}\vert f_{l,k}-f_{l,k+1}\vert ^{p}
\end{align*}
and the isotropic
\begin{align*}
\displaystyle \Vert f \Vert^{p}_{TV^{(\text{I})}}&=\sum_{l=1}^{m-1}\sum_{k=1}^{n-1}\sqrt{(f_{l,k}-f_{l+1,k})^{2p}+(f_{l,k}-f_{l,k+1})^{2p}}\\ &\ \ \ \ \ +\sum_{l=1}^{m-1}\vert f_{l,n}-f_{l+1,n}\vert ^{p}+\sum_{k=1}^{n-1}\vert f_{m,k}-f_{m,k+1}\vert ^{p}.
\end{align*}
See \cite{lasica2017total} for further details.
%-------------------------------------formulation---------------------------

%%%%%%%%%%%%%%%%%%%%%%%%%%%%%%%%%%%%%%%%%%%%%%%%%%
\section{Algorithm}
\label{sec:ALG}

The optimization problem defined in \eqref{eq:opt} can be solved via gradient-based methods. 
In particular, we use the second-order Taylor series expansion of $F(f)$ at the current iterate $f^j$ to define a sequence of quadratic sub-problems. The first and second derivatives of $F(f)$ can be computed exactly:
\begin{align*}
    \nabla F(f) &=\sum_{i=1}^m 
    \left ( \frac{r+y_i}{r+(Af)_i} - \frac{y_i}{(Af)_i}
    \right ) A^{\top} e_i\\
    \nabla^{2} F(f) &= A^{\top} \!  \left [ \sum_{i=1}^m 
    \left (
    \frac{r+y_i}{(r+(Af)_i)^2} -   \frac{y_i}{(Af)_i^2}
    \right )
    e_ie_{i}^{\top}
    \right ] \! A,
\end{align*}
where $e_i$  represents the $i^{\text{th}}$ column of the $m \times m$ identity matrix.

To reduce the computational complexity, we will apply the Barzilai-Borwein approach \cite{BarB88} to avoid computing the Hessian. 
Specifically, we approximate the second derivative of $F(f)$ with a scalar multiple of the identity matrix, i.e., $\nabla^2 F(f) \approx \alpha_j I$, where $\alpha_j > 0$ is a scalar. In the absence of the $\ell_p$ TV quasi-seminorm, the quadratic approximation is given by
$$
    F^j(f) = F(f^{j}) + (f - f^{j})^{\top}\nabla F(f^{j}) + \frac{\alpha_{j}}{2}\Vert f - f^{j}\Vert_{2}^{2}.
$$
By defining $q^j = f^{j} - \frac{1}{\alpha_{j}} \nabla F(f^{j})$ and incorporating the penalty term, the corresponding quadratic subproblems and iterates can be expressed as
\begin{align}\label{eq:SQP2}
    f^{j+1} \ = \  &  \ \underset{f \in \mathbb{R}^n}{\text{arg min}} \quad \ \ \  \frac{1}{2} \|f - q^{j} \|_{2}^{2} + \frac{\tau}{\alpha_j} \|f\|_{TV}^p \nonumber \\
    &\ \hbox{subject to} \quad  0 \leq f
\end{align}

To deal with the $\ell_p$ TV quasi-seminorm, we take the reweighted approach \cite{yan2015image, candes2008enhancing}, which allow us to transfer an $\ell_p$ TV nonconvex regularization directly to an $\ell_1$ TV convex regularization:
\begin{align*} 
\Vert f\Vert^p_{TV_{\gamma \omega}^{(\text{A})}}& = \displaystyle \sum_{l=1}^{m-1}\sum_{k=1}^{n}\gamma_{l,k}\vert f_{l,k}-f_{l+1,k}\vert \  + \\ 
&\ \ \ \ \  
\sum_{l=1}^{m}\sum_{k=1}^{n-1}\omega_{l,k}\vert f_{l,k}-f_{l,k+1}\vert, 
\end{align*}
and 
\begin{align*} 
&
\hspace{-.22cm} \Vert f\Vert^p_{TV_{\gamma \omega}^{(\text{I})}} =\\ &\displaystyle \sum_{l=1}^{m-1}\sum_{k=1}^{n-1}\sqrt{(\gamma_{l,k}(f_{l,k}-f_{l+1,k}))^{2}+(\omega_{l,k}(f_{l,k}-f_{l,k+1}))^{2}}\\ &+\displaystyle \sum_{l=1}^{m-1}\gamma_{l,n}\vert f_{l,n}-f_{l+1,n}\vert +\sum_{k=1}^{n-1}\omega_{m,k}\vert f_{m,k}-f_{m,k+1}\vert
\end{align*}
where the weights $\gamma_{l,k}$ and $\omega_{l,k}$ are defined as
\begin{align*} 
&\gamma_{l,k} = \vert f_{l,k}^{j}-f_{l+1,k}^{j}\vert^{(p-1)},\\ 
&\omega_{l,k} = \vert f_{l,k}^{j}-f_{l,k+1}^{j}\vert^{(p-1)},
\end{align*}
and $f^j$ is the output from the previous iteration $j$. 
These weights satisfy the following approximations: $$\vert f_{l,k}^{j}-f_{l+1,k}^{j}\vert^{(p-1)}(f_{l,k}-f_{l+1,k}) \approx (f_{l,k}-f_{l+1,k})^{p}$$ (see \cite{yan2015image} for details). 
We can now apply the fast gradient projection method \cite{beck2009fast} to solve the subproblem defined in \eqref{eq:SQP2}.

%----------------------------------------experiment ---------------------------
%----------------------------------------experiment ---------------------------
\section{Experiments}
\label{sec:EXP}
We performed two experiments with data using various noise levels, with dispersion parameters $r=1, 10, 25$, and 1000. All experiments involved reconstructing a blurry, noisy image (from the MATLAB Medical Imaging Toolbox) with dimensions of 512 by 512 pixels. \textbf{Each experiment was conducted 10 times, where the average of these trials are presented.} The objectives of the two experiments are:

\medskip

\begin{enumerate}[leftmargin=*,noitemsep,nolistsep]
\vspace{0.1cm}
\item To examine the effectiveness of the $\ell_p$ TV quasi-seminorm using different values of $p$.
\item To evaluate and compare the performance of the Poisson and negative binomial models when employing the $\ell_p$ TV quasi-seminorm.
\item To benchmark the performance of $\ell_p$ TV quasi-seminorm, $\ell_1$ TV quasi-norm, and $\ell_p$ quasi-norm.
\end{enumerate}
\vspace{0.2cm}

\smallskip
\noindent 
The details of each experiment are as follows:

\medskip

\noindent 
 \textbf{Experiment 1:} This experiment investigated the optimal solutions using negative binomial model with different $p$ values using the $\ell_p$ TV quasi-seminorm. The test data for this experiment consist of 2D images that have different noise-level intensities, drawn from the negative binomial distribution.

 \smallskip
 
%\item 
\noindent 
 \textbf{Experiment 2:} Here, we compared the results from using Poisson and negative binomial models. This experiment focuses on the relative performances of multiple types of regularization, including the $\ell_p$ TV quasi-seminorm, $\ell_1$ TV quasi-norm, and $\ell_p$ quasi-norm. As with the first experiment, the test data were the same 2D images with noise drawn from a negative binomial distribution.
%\end{itemize}

For the negative binomial models, several methods exist for estimating the parameter $r$, such as method-of-moment \cite{clark1989estimation} and the maximum quasi-likelihood methods \cite{piegorsch1990maximum}. Additionally, cross-validation techniques offer a precise method for estimating the dispersion parameter \cite{gu1992cross}. However, for the sake of eliminating potential biases or inaccuracies inherent in the parameter estimation process, our experiments intentionally use the exact value of the parameter $r$ as a prior.
%----------------------------------------Experiment I------------------------
\begin{figure}[!b]
\includegraphics[trim=4.5cm 0 0 0,clip,width=9.25cm]{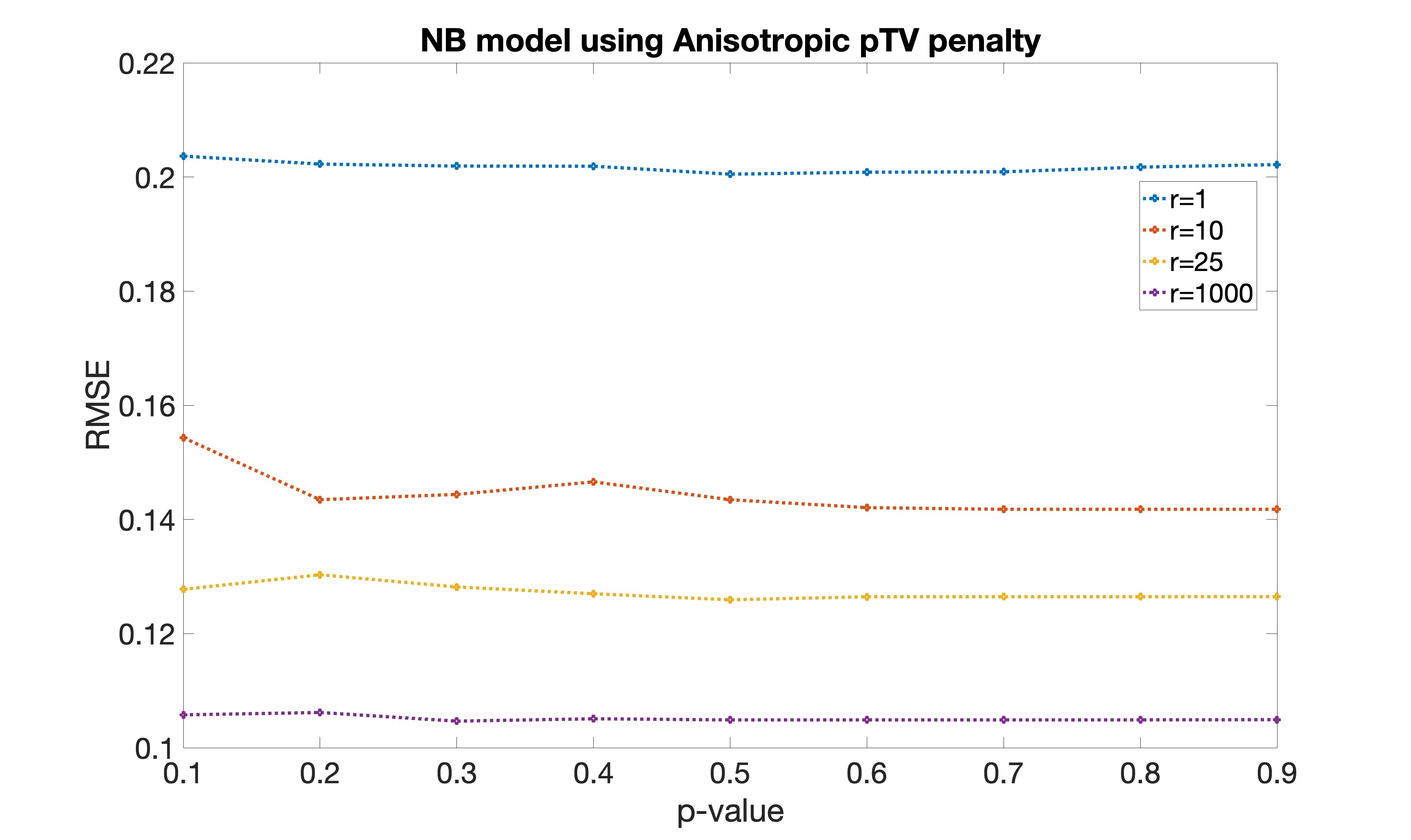}
\includegraphics[trim=4.5cm 0 0 0,clip,width=9.25cm]{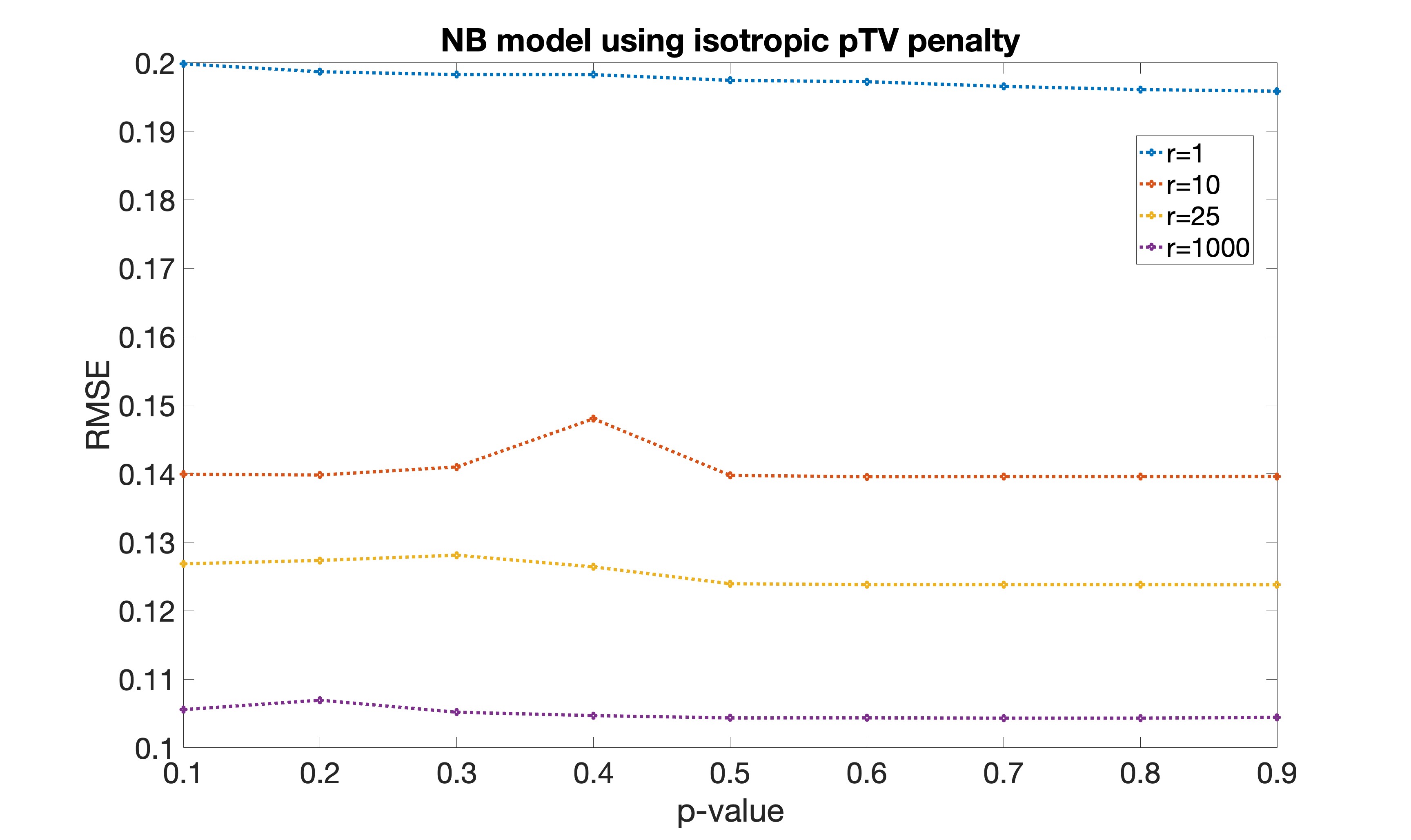}

\caption{\textbf{Experiment I:} RMSE analysis for 2D data reconstruction employing the negative binomial model in conjunction with the $\ell_p$ TV quasi-seminorm. The evaluation spans across multiple noise levels, i.e., $r=1, 10, 25$ and $1,000$ for different $p$ values. Observe that the RMSE values do not change significantly as the value of $p$ changes.}
\label{fig:Exp1}
\end{figure}

\subsection{Experiment I: $p$-value exploration}
In our first experiment, we examined a set of nine distinct $p$ values, evenly spaced between 0.1 and 0.9, using the negative binomial model with both anisotropic and isotropic $\ell_p$ TV quasi-seminorms. These were applied to four different noise-level images corresponding to four different dispersion parameter $r$ values. As expected, as the parameter $r$ increased, the corresponding RMSE decreased. When $r=1$, the RMSE was highest. In contrast, the scenarios with the highest $r$ value (specifically $r=1000$) returned the lowest RMSE.

Interestingly, our findings indicate that the results of both isotropic and anisotropic $\ell_p$ TV quasi-seminorms are not sensitive to variations in the $p$ value, as shown in Fig. \ref{fig:Exp1}. Specifically, across the range of $p$ values, the RMSE values change by less than 1\%. 
% This observation contrasts with the behavior of the $\ell_p$ quasi-norm, where RMSE changes spanned up to 7\% \cite{LuMpNorm23}.

%----------------------------------------Experiment II-----------------------
\subsection{Experiment II: Comparison between Poisson and negative binomial models with mutiple regularizations}
\begin{figure}[t!]
\begin{tabular}{cccc}
&
Poisson & Negative Binomial \\[-.1cm]
&
Reconstruction & Reconstruction \\
{\rotatebox[origin=c]{90}{
\hspace{3.5cm} 
$\ell_p$ TV-A norm
}} 
\hspace{-.22cm}
&
\hspace{-.26cm} 
\includegraphics[width=3.6cm]{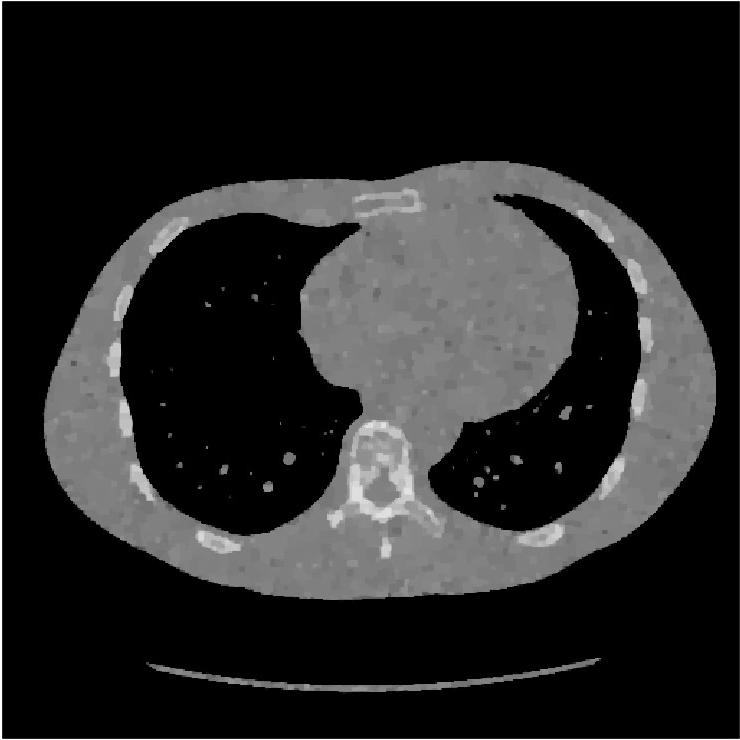} 
\hspace{-.22cm} 
&
\hspace{-.22cm} 
\includegraphics[width=3.6cm]{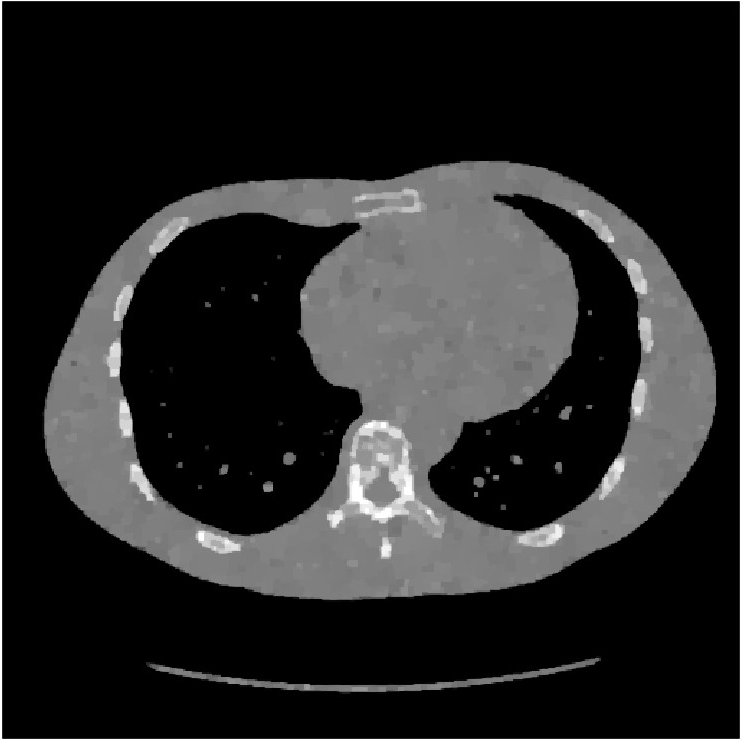} 
\hspace{-.22cm}  &  \ \ 
\\[-2.7cm]
& (a) RMSE = 15.35\% & (b) RMSE = 13.96\%  \\[.2cm]
{\rotatebox[origin=c]{90}{
\hspace{3.5cm} $\ell_p$ TV-I norm
}} 
\hspace{-.22cm}
&
\hspace{-.26cm} 
\includegraphics[width=3.6cm]{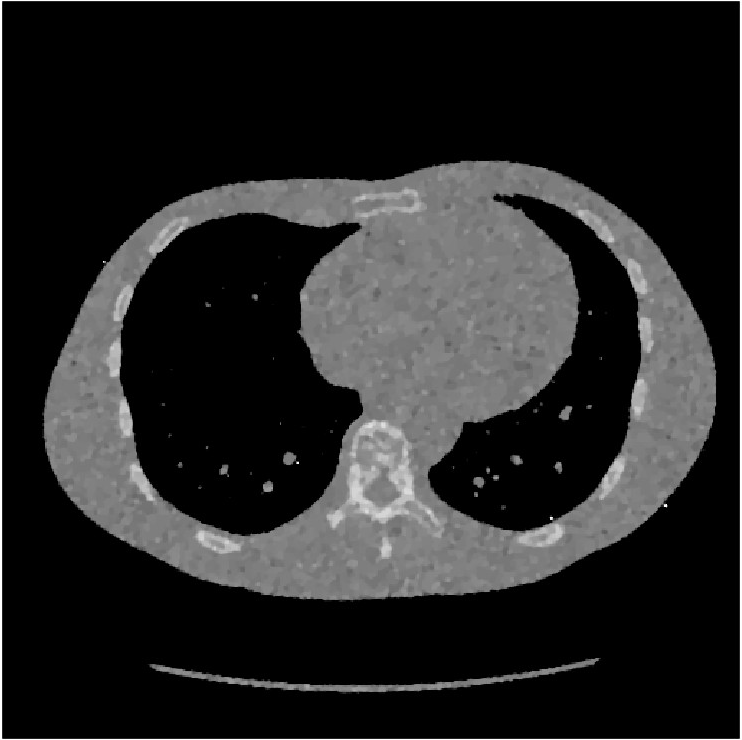} 
\hspace{-.22cm} 
&
\hspace{-.22cm} 
\includegraphics[width=3.63cm]{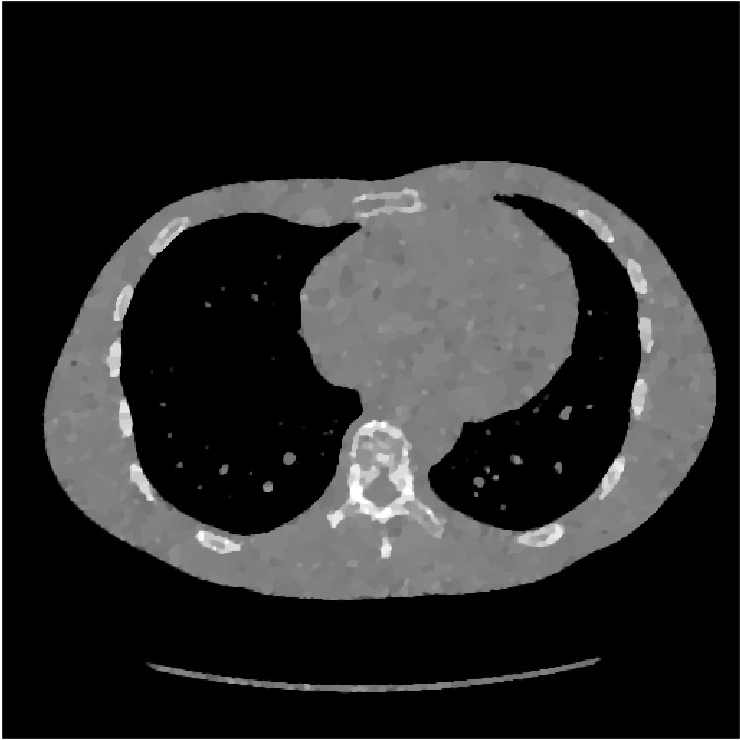} 
\hspace{-.22cm}  & 
\\[-2.7cm]
& (c) RMSE = 15.36\% & (d) RMSE = 14.18\% 
\end{tabular}
\caption{\textbf{Experiment II:} 2D data from a negative binomial distribution ($r=10$). The negative binomial model yields a lower RMSE than the Poisson model with $\ell_p$ TV quasi-seminorm. The difference between isotropic and anisotropic versions is small.}
\label{fig:Exp2}
\end{figure}

In the second experiment, we conducted a comparative analysis between the negative binomial and Poisson models using three different penalties. The results for these comparisons are presented in Table \ref{table:RMSE}. Additionally, the results of $r=10$ are visually represented in Fig.\ \ref{fig:Exp2}.

Our observations consistently show that the negative binomial model achieve lower RMSE values than the Poisson model with various $r$ values. As $r$ increases, the RMSE gaps between the two models diminish (see Table \ref{table:RMSE}). This aligns with the fact that the Poisson distribution can be seen as a limiting case of the negative binomial distribution when the dispersion parameter $r$ approaches infinity.

\begin{center}
\begin{table}[!h]
\centering
\begin{tabular}{|l|l|c|c|c|c|}
\hline
\multirow{2}{*}{Model}                    
& \multirow{2}{*}{\ Penalty}
& \multicolumn{4}{c|}{Dispersion parameter $r$} \\ \cline{3-6}
& 
& $1$ & $10$ & $25$ & $1,000$ \\ \hline      
\parbox[t]{2mm}{\multirow{5}{*}{\rotatebox[origin=c]{90}{
\begin{tabular}{c}Negative\\Binomial
\end{tabular}}}}
& $\ell_p$ TV-A & 0.2005     & \textbf{0.1396}     & \textbf{0.1238}      & 0.1047       \\ \cline{2-6} 
& $\ell_p$ TV-I & \textbf{0.1957}    & 0.1418      & 0.1259      & \textbf{0.1043}       \\ \cline{2-6} 
& $\ell_1$ TV-A & 0.2019    & 0.1421     & 0.1267      & 0.1053        \\ \cline{2-6} 
& $\ell_1$ TV-I & 0.1959    & 0.1399     & 0.1240     & 0.1047       \\ \cline{2-6} 
& $p$ norm       & 0.2104    & 0.1603      & 0.1411     & 0.1253       \\ \hline
\hspace{.2cm}  \parbox[t]{2mm}{\multirow{5}{*}{\rotatebox[origin=c]{90}{Poisson}}} 
& $\ell_p$ TV-A & 0.2364     & 0.1535      & 0.1403      & 0.1054       \\ \cline{2-6} 
& $\ell_p$ TV-I & 0.2231    & 0.1536     & 0.1422      & 0.1053       \\ \cline{2-6} 
& $\ell_1$ TV-A & 0.2386    & 0.1569     & 0.1501     & 0.1062       \\ \cline{2-6} 
& $\ell_1$ TV-I & 0.2253    & 0.1581     & 0.1484     & 0.1054       \\ \cline{2-6} 
& $p$ norm      & 0.2391    & 0.1739     & 0.1526      & 0.1262       \\ \hline
\end{tabular}
\caption{\textbf{Experiment II:} 
Evaluation of the Root-Mean-Square Error (RMSE) for both the negative binomial and Poisson models with four dispersion parameters ($r = 1, 10, 25$, and $1,000$) using five regularization techniques: isotropic $\ell_p$ TV ($\ell_p$ TV-I), anisotropic $\ell_p$ TV ($\ell_p$ TV-A), isotropic $\ell_1$ TV ($\ell_1$ TV-I), anisotropic $\ell_1$ TV ($\ell_1$ TV-A), and the $p$ quasi-norm. For each penalty, the parameter $p$ is optimized to yield the minimum RMSE.
}\label{table:RMSE}
\end{table}
\end{center}

\vspace{-0.5 cm}
Furthermore, our analysis reveals that the RMSE differences between anisotropic and isotropic $\ell_p$ TV quasi-seminorms are consistently small, regardless of whether the negative binomial or Poisson model is applied. Additionally, $\ell_p$ TV quasi-seminorms consistently achieve lower RMSE values than $\ell_p$ quasi-norms. We generally observe an improvement in RMSE values using the $\ell_p$ TV quasi-seminorm over the $\ell_1$ TV quasi-seminorm.

%Given that the $\ell_1$ TV quasi-seminorm can be considered a special case of the $\ell_p$ TV quasi-seminorm where $p=1$, this result aligns with the finding of Experiment I.

%\FloatBarrier

%----------------------------------------conclusion------------------------
\section{Conclusion}
In this study, we investigate the $\ell_p$ TV quasi-seminorm for signal reconstruction using 2-D data drawn from a negative binomial model with four different noise levels. Our findings demonstrate that the performance of the $\ell_p$ TV quasi-seminorm surpasses that of the $\ell_p$ quasi-seminorm in both negative binomial and Poisson model. Furthermore, our results show that the $\ell_p$ TV quasi-seminorm's performance remains consistent when using different $p$ values within the range $0 < p \leq 1$. Additionally, when comparing the negative binomial model with the Poisson model, our findings verify that the negative binomial model yields lower RMSE values in signal reconstruction.

% Below is an example of how to insert images. Delete the ``\vspace'' line,
% uncomment the preceding line ``\centerline...'' and replace ``imageX.ps''
% with a suitable PostScript file name.
% -------------------------------------------------------------------------

%\vfill
%\pagebreak

\newpage

\bibliographystyle{IEEEbib}
\bibliography{ref}

\end{document}